\documentclass[aps,pre,showpacs]{revtex4}
\usepackage{epsfig}
\def\dt{\Delta t}
\begin{document}
\title{Time dependent cross correlations between different stock
returns: A directed network of influence}
\date{\today}
\author{L. Kullmann$^1$, J. Kert\'esz$^{1,2}$, K. Kaski$^2$}
\affiliation{$^1$Department of Theoretical Physics, Budapest University of
Technology and Economics, Budafoki \'ut 8, H-1111, Budapest, Hungary}
\affiliation{$^2$ Laboratory of Computational Engineering, Helsinki University
of Technology, P.O.Box 9400, FIN-02015 HUT, Finland}
\begin{abstract}
We study the time dependent cross correlations of stock returns,
i.e. we measure the correlation as the function of the time shift
between pairs of stock return time series using tick-by-tick data. We
find a weak but significant effect showing that in many cases the
maximum correlation appears at nonzero time shift indicating
directions of influence between the companies. Due to the weakness of
this effect and the shortness of the characteristic time (of the order
of a few minutes) our findings are compatible with market
efficiency. The interaction of companies defines a directed network of
influence.
\end{abstract}
\maketitle
\section{Introduction}
In the risk minimisation of portfolio optimisation it is very
important to consider how the returns of different companies correlate
with each other. For this purpose the study of the equal time cross
correlations between stocks has attracted much
interest~\cite{MANT,KLKJ}: The clustering properties and the
comparison between the time and ensemble averages have provided much
useful information in this respect.

In statistical physics {\it time dependent} correlations are also of
major interest. Due to their role in the fluctuation dissipation
theorem they constitute as the main tool for determining transport
coefficients. The famous Onsager reciprocity relations have their
roots in the symmetry properties of the time-dependent cross
correlations~\cite{STATFIZ}.  Obviously, in an economic system there
is no reason to assume that the time reversal symmetry or detailed
balance is maintained.  Nevertheless, it is of interest to investigate
the time dependent cross correlations between stock returns because
they contain information about the way how the prices influence each
other, which are the dominant stocks and to what extent this dominance
is reflected in the price changes under the conditions of an efficient
market.

The time dependent correlations between the indices of different stock
exchanges were already studied empirically by \cite{LIN,VANDEW} and
with a microscopic model by \cite{SCHULZE}. They showed that there
exists a time shift in the cross-correlations which arises from the
fact that the different stock markets are open in different time
cycles during the day as the Earth rotates.

In this article we study time-dependent cross correlation functions of
the returns of different stocks taken from the New York Stock Exchange
(NYSE). As we will show, in many cases the maximum of the correlation
as a function of time is not at zero but shifted, meaning that there
exists some ``pulling'' effect between the companies, {\em i.e.,} one
of them influences the price behaviour of the others. However, this
effect cannot be strong and the shift should be small, otherwise the
effect could be utilized for arbitrage purposes, which is excluded
from an efficient market. In fact, the investigated cases do not
contradict with these criteria. The time dependent correlation between
the stocks was studied before by \cite{LO,CAMBELL}. The results of
\cite{LO} seems to contradict with ours because they studied weakly
returns, and they found significant cross-correlations on the weakly
scale. However, the results agree considering the ``pulling'' effect,
namely that the cross-correlations are asymmetric.

The paper is organized as follows: In the next Section we give a short 
description of the data set. In Section 3 we present the method of analysis
and show how it works on an artificial set of data. Section 4 is
devoted to the presentation of the results. The paper terminates with
a discussion. 

\section{Data}
\label{Data}
One of the stylized facts of markets is that the auto-correlation of
stock returns decays exponentially with a very short characteristic
correlation time, which is in the range of a few minutes
\cite{MS,LIU}. This is understood as a signature of market
efficiency\cite{FAMA}. Since cross correlations could also be used for
arbitrage, one should not expect effects much beyond the above scale
and therefore high frequency data are needed. We have analyzed the Trade
and Quote (TAQ) database for $N=54$ days over the time period from
01.12.1997 to 09.03.1998, which includes tick-by-tick data for about
10000 companies. Since this is quite a short time period we selected
only those companies which were traded more than 15000 times such that
the number of companies reduced to 195.

Having these 195 time series we have to face the following problem:
since the tradings do not happen simultaneously, the values of the
returns have also to be defined for the time intervals between the
tradings.  According to the rules of the stock exchanges we have
considered the price as constant between two changes. The whole
trading time $T$ during one day is divided into $n$ small intervals or
windows of size $\Delta t = T/n$. If the trading happens in the
interval $t$ the return takes the value
\begin{eqnarray}
r_{\Delta t}(t) = {\ln\left[ p(t) \right] 
\over \ln\left[ p(t-\Delta t) \right]}\  \nonumber, 
\end{eqnarray}
where for simplicity the day index $i$ is not indicated; otherwise it
is zero.

In order to avoid the problem of major return values stemming from the
differences between opening and previous day's closing prices we
simply took the days as independent, {\it i.e.}, the averaging is
separated into two steps: Over the intraday trading time $T$ and over
the trading days.
The data prepared in this way were then analysed from the point of view
of time-dependent cross correlations.

\section{Method of measuring the correlation}
\label{measuring_correl}
As mentioned in the introduction we want to investigate the correlation 
of returns as the function of the time shift between pairs of
stocks' return time series. The definition of the time-dependent
correlation function $C_{A,B}(\tau)$ is
\begin{equation}
\label{correl_def}
C_{\dt}^{A,B}(\tau) = {\langle r_{\dt}^A(t) r_{\dt}^B(t+\tau) \rangle -
\langle r_{\dt}^A(t) \rangle \langle r_{\dt}^B(t+\tau) \rangle \over
\sigma_A \ \sigma_B}\ ,
\end{equation}
where $\sigma^2 = \left\langle \left( r_{\dt}(t) - \langle r_{\dt}(t)
\rangle \right)^2 \right\rangle$ is the variance of the return. The
notation $\langle . \rangle$ means averaging over the whole trading
time $T$ and important details of this process will be given in the
following.

Since the smallest interval between two tradings is one second, then 
$\dt = 1 s$ seems to be a natural choice. However, for such a short window 
it quite often happens that at a given time step there is no transaction
for one of the stocks (or for both) such that the return results in a zero
contribution to the total correlations.  Since the number of non-zero
contributions is small, the correlation coefficients as a function of
the time shift, $\tau$, will strongly fluctuate. To avoid this problem
one has to enlarge the time difference, $\dt$, and average the
correlations over the starting points of the returns. 
In this way the average in Eq. (\ref{correl_def}) means the following
\begin{equation}
\label{correl_average}
\left\langle r_{\dt}^A(t)\ r_{\dt}^B(t+\tau) \right\rangle = 
{1 \over T} \ \sum_{t_0 = 0}^{\dt-1} \ \sum_{k=1}^{T/\dt}
r_{\dt}^A(t_0 + k \dt) r_{\dt}^B(t_0 + k \dt + \tau) \ ,
\end{equation}
where the first sum runs over the starting points of the returns and
the second one runs over the $\dt$ wide windows of the returns.

In order to illustrate the effect that by taking larger time
difference it is easier to identify the peaks in the correlation
function -- in other words to locate the time which gives the maximal
correlation -- we simulated two series of artificial data sets. The
first is a one dimensional Persistent Random Walk (PRW)~\cite{FURTH},
which deviates from a normal RW by the fact that the probability,
$\alpha$, that it jumps in the same direction as in the previous step
is higher than $0.5$, {\it i.e.}, the random walker remembers its
history. The probability of an increment $x(t)\in\{\pm 1\}$ at time
$t$ is
\begin{equation}
P \left( x(t) \right) = 
\alpha \delta_{x,x(t-1)} + (1-\alpha) (1-\delta_{x,x(t-1)}).
\end{equation}
The other time process is simply generated from the first one by
shifting it by $\tau_0$ and adding to it Gaussian random noise with 
zero mean and width $\sigma $:
\begin{equation}
y(t) = x(t-\tau_0) + \xi(t), \ \ \ \xi \in N(0,\sigma)
\end{equation}
The advantage of this model is that the correlation function can be
calculated analytically and the position of the maximum
correlations can be adjusted at $\tau_0$:
\begin{equation}
\label{anal_artific}
C(\tau) = {(2\alpha - 1)^{|\tau-\tau_0|} \over
\sqrt{\sigma+1}}
\end{equation}
After generating the two data sets we randomly drop points from both sets
and keep only the fraction $\rho$ of the points in order to have the
same problem as with the original data sets that the jumps do not occur
at the same time in the different time series.
\begin{figure}[ht]
\centerline{\epsfig{file=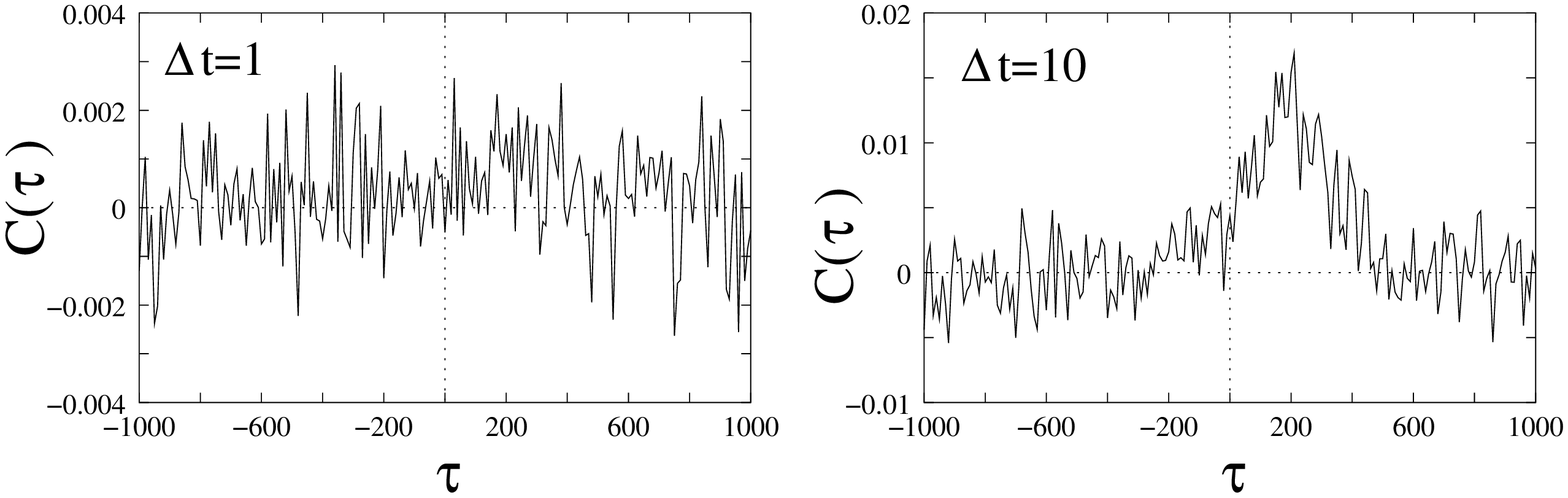,width=5.2in}}
\caption{Illustration how the correlation as the function of the time
shift, $\tau$ depends on the time deference, $\dt$ of the return. The
correlation was measured on the artificial data sets with parameters:
$\rho = 0.01, \tau_0 = 200, \sigma = 1000, \alpha = 0.99.$
The figures shows that while for $\dt=1$ no peak in the
correlation function can be identified, by increasing the time
difference to $\dt=10$ the peak at $\tau = 200$ appears.}
\label{artificial}
\end{figure}
It is apparent from Fig.\ref{artificial}. that increasing the time
difference, $\dt$ helps identifying the time of maximum
correlation. 
The fact that we dropped random points from the original data
changes slightly the position of the maximum correlation as compared
to Eq.(\ref{anal_artific}).

Fig.\ref{artificial}. shows that the decay of the correlation function
is not exponential as in Eq.(\ref{anal_artific}) but it decays
approximately linearly down to the noise level. This is due to the
averaging procedure we use with the increased time difference, $\dt$.
The correlation corresponding to larger time difference
$C_{\dt}(\tau)$ can be written as the weighted sum of the one-step
correlation functions, $C_1$, which belongs to $\dt=1 s$:
\begin{eqnarray}
\label{correl_atiras}
\left\langle r_{\dt}^A(t) r_{\dt}^B(t+\tau) \right\rangle &=&
\left\langle \sum_{i=1}^{\dt} \delta^A(t+i) \sum_{i=1}^{\dt}
\delta^B(t+i+\tau) \right\rangle = \\
&=& 1 C_1(\tau-\dt+1) + \dots
+ (\dt-1)\ C_1(\tau-1) + \nonumber \\
&& +\ \dt\ C_1(\tau) + \dots + 1 C_1(\tau+\dt-1)\ , \nonumber
\end{eqnarray}
where $\delta = r_{\dt = 1}$ is the return belonging to one second
time difference.

Changing $\tau$ in Eq.(\ref{correl_atiras}) means changing the weights
of the one-step correlation functions. Since the correlation function
of the original data sets, see Eq.(\ref{anal_artific}), decays
exponentially, the maximum, $C_1(\tau_0)$, will give the main
contribution to the sum in Eq.(\ref{correl_atiras}) and because its
weight is linear in $\tau$, then $C_{\dt}(\tau)$ will decay approximately
linearly. (It should be noted that the normalization factor in 
Eq.(\ref{correl_def}) does not change this consideration since it 
is independent of $\tau$.)

There is only one question left namely how can we choose a smaller
value for $\tau$ than for $\dt$? The time dependent cross correlation
of the returns contains a product of the {\it return} of company A
with that of company B shifted by $\tau$. As the return is defined
with the window $\dt $ the values of $\tau $ could only be multiples
of $\dt$. The solution is simply that one shifts the starting point of
the return of company B by $\tau$, as evident in
Eq.~(\ref{correl_average}), i.e.  we make the time shift in the
price function and in this way allow any time shift larger than the
minimum trading time.

The above arguments of averaging give support to choose a value 
for $\dt$ that is larger than the minimum trading time. However, 
it should not be too large since the averaging leads then to the 
smearing out of the maximum. As the width of the one-step correlation 
should be a few minutes, much larger time difference would mean 
that in the sum of Eq.(\ref{correl_atiras}) we mainly have terms, 
which are only due to noise. This suggests that the optimal choice 
for $\dt$ is of the order of magnitude of one minute.

\section{Results}
As mentioned in Sec. \ref{Data} we have studied the correlation of 195
companies, which were traded during the available 54 days more than
15000 times. In accordance with the arguments presented in the
previous Section we have used $\dt=100$ but checked that the results
are quite robust within the range $50 \le \dt \le 500$. As already
mentioned, we averaged over the starting points of the returns. For
the maximum of the time shift we choose $2000 s$. This is definitely
beyond any reasonable characteristic time for correlations in return
values because of market efficiency. In fact, using such a large value
for the time shift allows us to measure the noise level, which the
possible effect should be compared with.

\begin{figure}
\centerline{
\epsfig{file=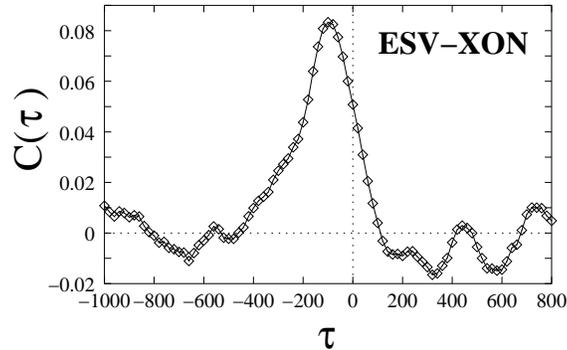,width=3in}}
\caption{
Example for the measured shifted-time correlation function. The two
companies are: Ensco International (ESV) and Exxon Corp. (XON). The
maximum correlation value is at -100 sec, which means that the return
time series of ESV has to be shifted back in order to get the maximal
correlation i.e. the price changes happen later in time, in other words 
ESV is pulled by XON.
}
\label{esv-xon}
\end{figure}

For the resulting 195*194/2 correlation functions we measured the
maximum value, $C_{max}$, the position $\tau_{max}$ at which time
shift this maximum was found, and the ratio, $R$, of the maximum and
strength of the noise defined as the variance of the
correlation values for time shift values between $600$ and
$2000s$. We looked at those pairs of companies for which these three
values exceeded a prescribed threshold values, which we defined for
$\dt = 100$ as: $\tau_{max} \ge 100$, $C_{max} \ge 0.04$, $R \ge 6.0$.
One example of the measured correlation function can be seen in
Fig.~\ref{esv-xon}. In this case the company XON (Exxon) -- which is a
large oil company -- "pulls" the ESV (Ensco International) which
provides drilling service to oil and gas companies. This effect is
quite weak but the large value of $R$ shows that it is significant.

The maximal value of the correlations turn out to be quite small, in
average less than 0.1, ({\it e.g.} see Fig. \ref{esv-xon}), although
the generally quoted equal time cross correlations have much larger
values. The root of this effect lies in the choice of the time
difference, $\dt$. Increasing $\dt$ increases the values of the equal
time correlations \cite{MANT2}.

In some cases the position of the maximum correlation was found at
values much larger than few minutes which would be inconsistent with
the efficient market behavior. A closer inspection revealed that in
such cases the peak in the correlation function is caused by two major
return values in the considered time series. The contribution of their
product to the correlation -- at appropriate value of the time shift
-- dominates the maximum of the correlation function. These are not
the effects we are looking for, therefore we did not take them into
account. In order to check whether the peak in the correlation is due
to some single large return value or due to persistent influence of
one of the stocks on the other we also studied how the correlation
changes if the analyzed time window changes. We measured the shifted
time correlation also for the first and for the second half of the
given 3 month period and studied whether the correlation function
remains qualitatively the same.

\begin{figure}[ht]
\centerline{
\epsfig{file=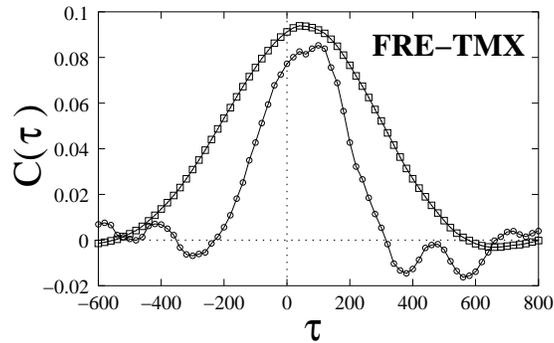,width=3in}}
\caption{Example for a pair of company for which the correlation
function has an asymmetric peak. The curve with circles belongs to
$\dt = 100$, the other with the squares to $\dt = 500$. The
maximum of the second curve is at smaller time value because the left side
of the peak -- in the case of the curve with circles -- is higher.}
\label{fre-tmx}
\end{figure}

We also measured the correlation for shorter and for larger time
difference, {\it i.e.} $\dt=50$ and $\dt=200$, respectively, because 
it may happen that by changing the
time difference also the position of the maximal correlation value
changes due to the averaging procedure described in
Eq.(\ref{correl_atiras}).  This can happen if the time dependent
correlation function for $\dt$ has an asymmetric peak; see
Fig.~\ref{fre-tmx}. Let us suppose that the left hand side is higher 
than the right hand one. For $\dt' > \dt$ the maximum will be 
shifted towards left as it can be shown through simple examples using
Eq.(\ref{correl_atiras}). In the case of Fig.~\ref{esv-xon} the
correlation function is also asymmetric but not at its peak (not near
the maximum), which means that the maximum will not be shifted by
increasing the time difference, $\dt$.

The results show that the characteristic time shift is around 100
sec. which is consistent with the effective market hypothesis. A time
shift larger than the characteristic time of the decay of the
return auto-correlations would contradict with the efficient market
picture and could be used to arbitrage.

In general the more frequently traded companies are influencing 
(``pulling'') the less frequently traded ones. This is not
surprising since obviously the more frequently traded companies are
more important. It is therefore more likely that they influence a
smaller company than the other way around.  Although this is the
generic situation, there are a few exceptions when a less often traded
company ``pulls'' the other one.

\begin{figure}
\centerline{
\epsfig{file=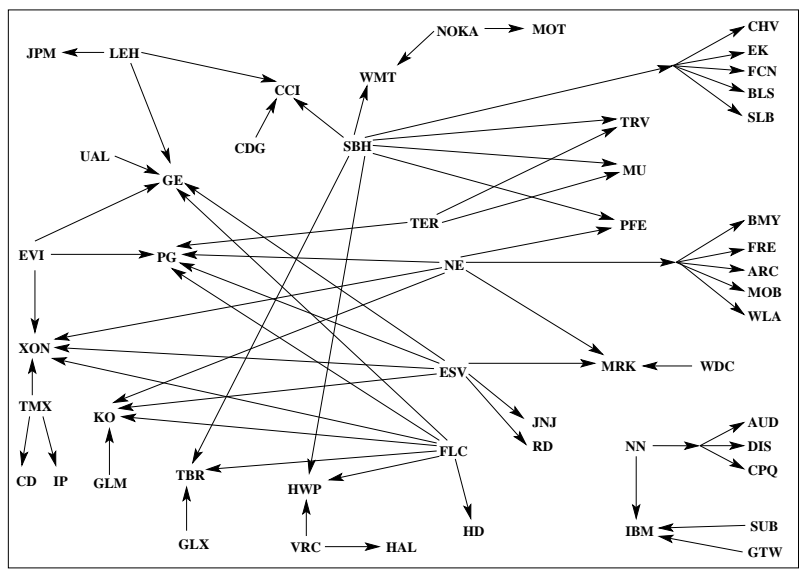,width=300pt,height=212pt}}
\caption{
Representation of the pulling effect between the companies. The 
direction of the arrows show which company is pulling the other. The 
companies which appear in the figure show the most significant effects.
}
\label{arrow}
\end{figure}

In this study we found that in general one ``small'' company is
influenced by many ``large'' companies and one ``large'' company pulls
many ``small'' ones. As can be seen in Fig. \ref{arrow} this behaviour
can be represented as a graph of directed links, where there are nodes
from which many links go out (meaning that this node is influenced by
many others) and there are other nodes where many links go in (these
are the big companies influencing the less important ones).

\section{Discussion}
In this paper we have analyzed the time dependent cross correlation
functions of the returns of stocks at the NYSE. We have studied whether 
there exists any pulling effect between stocks, i.e., whether at a given 
time the return value of one stock influences that of another stock at
a different time.

In general we can see two types of mechanisms to generate significant
correlation between two stocks:
\begin{itemize}
\item[(i)] Some external effect (e.g. economic, political news etc.) that
influences both stock prices simultaneously. In this case the change
for both prices appears at the same time, and the maximum of the
correlation is at zero time shift.
\item[(ii)] One of the companies has an influence to the other (e.g. one of
the company's operation depends on the other.). In this case the price
change of the influenced stock appears later in time because it needs
some time to react on the price change of the first stock, in other words
one of the stocks pulls the other. This pulling effect has been the main
focus of our study in this article.
\end{itemize}

Since the correlation between stocks was expected to be small and the
available set of data was somewhat limited we had to do a careful
analysis. For this reason and test purposes we generated an artificial
data set with which we showed that by increasing the time window of
the returns and by averaging over their starting points the detection
of the correlation effect gets easier.

With the real data we saw that it is possible to find pairs of stocks
where the pulling effect exists, though it turned out to be small. In
addition, the characteristic time shift - given by the position of
maximum correlation - was found to be of the order of a few minutes.
These findings are compatible with the efficient market picture.

As for the pulling effect we found that generically the more traded,
and thus more important companies pull the relatively smaller
companies. This result is consistent with that of \cite{LO}. In this
light it is not surprising that in the study of the time-dependent
cross correlation functions of pairs of companies in the Dow Jones
Industrial Average index no pulling effect was found. This underlines
the fact that the Dow Jones companies are indeed among the most
important stocks of the New York Stock Exchange.

Finally we would like to propose that although the observed pulling
effect was small, our careful analysis could show that it is
significant for a considerable set of pairs of companies. We think
that this property of the stock market should be added to the so
called stylized facts. Of course, further analysis on more extensive
data is needed to clarify further details of the time dependent cross
correlations.

{\bf Acknowledgment:} We are grateful to A. L. Barab\'asi for
discussions, to Soon-Hyung Yook and Hawoong Jeong for their help in
processing the data. Support by OTKA (T029985) and the Academy of
Finland, project No. 1169043 (Finnish Centre of Excellence Programme
2000-2005) are gratefully acknowledged.


\appendix
\section{Company names and description}
\begin{tabbing}
xxxxxxxxxx\=xxxxxxxxxxxxxxxxxxxxxxxxxxxxxxxxxx\=\kill
{\bf Symbol} \> {\bf Name} \> {\bf Description}\\
ARC  \> Atlantic Richfield Co.         \> Petroleum refining \\
AUD  \> Automatic Data Processing      \> Data communications and
information services. \\
BLS  \> Bellsouth Corp.                \> Telephone communication\\
BMY  \> Bristol-Myers Squibb Co.       \>  Pharmaceutical preparations \\
CCI  \> Citicorp                       \> Banking. \\
CD   \> Cendant Corp.                  \> Travel, real estate, vehicle,
and financial services\\ 
CDG  \> Cliffs Drilling Co.            \> International drilling company\\
CHV  \> Chevron Corp.                  \> Energy and chemical company,
petroleum refining\\
CPQ  \> Compaq Computer Corp.          \> Electronic computers\\
DIS  \> Walt Disney Co.                \> Entertainment company\\ 
EK   \> Eastman Kodak Co.              \> Photography \\
ESV  \> Ensco International Inc.       \> Drilling oil and gas wells \\
EVI  \> Energy Ventures Inc.           \> Oil and gas field machinery \\
FCN  \> First Chicago NBD Corp.        \> Banking \\
FLC  \> Falcon Drilling Co. Inc.       \> Marine-based drilling\\
GE   \> General Electric Co.           \> Electronics, machinery\\
GLM  \> Global Marine Inc.             \> Drilling oil and gas wells \\
GLX  \> Glaxo Wellcome Plc.            \> Pharmaceutical preparations \\
GTW  \> Gateway 2000 Inc.              \> Electronic computers\\
HAL  \> Halliburton Co.                \> Oil field services\\
HD   \> Home Depot Inc.                \> Home improvement retailer\\
HWP  \> Hewlett-Packard Co.            \> Computers\\
IBM  \> International Business Machines Corp.   \> Computers \\
IP   \> International Paper Co.        \> Paper \\
JNJ  \> Johnson \& Johnson             \> Health care products\\
JPM  \> Morgan J.P. Co. Inc.           \> Banking\\
KO   \> Coca-Cola Co.                  \> Soft drinks\\
LEH  \> Lehman Brothers Holdings       \> Financial services\\
MOB  \> Mobil Corp.                    \> Petroleum refining\\ 
MOT  \> Motorola Inc.                  \> Semiconductor technology\\ 
MRK  \> Merck \& Co Inc.               \> Pharmaceutical preparations\\
MU   \> Micron Technology Inc.         \> Semiconductor technology\\
NE   \> Noble Drilling Corp.           \> Drilling oil and gas wells\\
NN   \> Newbridge Networks Corp.       \> Telephone and telegraph apparatus\\
NOKA \> Nokia Corp.                    \> Mobile phones\\
PFE  \> Pfizer Inc.                    \> Pharmaceutical preparations\\
PG   \> Procter \& Gamble Co.          \> Soap and other detergents\\
RD   \> Royal Dutch Petroleum Comp.    \> Petroleum Refining\\ 
SBH  \> Smithkline Plc                 \> Pharmaceutical preparations\\
SLB  \> Schlumberger Limited LTD       \> Oil and gas field services\\
SUB  \> Summit Bank Corp.              \> Banking\\
TBR  \> Telecomunicacoes Brasileiras S.A.\> Telecommunications\\
TER  \> Teradyne Inc.                  \> Electrical instruments\\
TMX  \> Telefonos de Mexico            \> Telephone communication\\
TRV  \> Travelers Group Inc.           \> Fire, marine and casualty insurance\\
UAL  \> UAL Corp.                      \> Air transportation\\
VRC  \> Varco International Inc.       \> Oil and gas field services\\
WDC  \> Western Digital Corp.          \> Computer storage devices\\
WLA  \> Warner Lambert Co.             \> Pharmaceutical preparations\\
WMT  \> Wal-Mart Stores Inc.           \> Retail - Variety stores\\
XON  \> Exxon Corp.                    \> Petroleum refining\\
\end{tabbing}
\end{document}